\documentclass[preprint,authoryear,12pt]{elsarticle}%

\usepackage{amssymb}
\usepackage{graphicx}
\usepackage{epsfig}
\usepackage{amsthm}
\usepackage{longtable}
\usepackage{lscape}

\journal{New Astronomy Review}

\begin{document}

\begin{frontmatter}



\title{Basic Parameters of Open Star Clusters \astrobj{DOLIDZE~14} and \astrobj{NGC~110} in Infrared bands}


\author[ku]{Gireesh C. Joshi}
\ead{gireesh@aries.res.in}
\author[ari]{Y. C. Joshi}
\author[ari]{S. Joshi}
\author[kh]{R. K. Tyagi}
\address[ku]{Center of Advanced Study, Department of Physics, Kumaun University, Nainital-263002}
\def\astrobj#1{#1}
\address[ari]{Aryabhatta Research Institute of Observational Sciences, Nainital, India 263002}
\address[kh]{Department of Physics, Hemwati Nandan Bahuguna Govt. P.G. College, Khatima-262308, Uttarakhand}
\begin{abstract}
The basic physical parameters of a poorly studied open cluster \astrobj{NGC~110} and an unstudied open cluster \astrobj{DOLIDZE~14} are estimated in the present study using the archival $PPMXL$ and $WISE$ catalogues. The radius of both the clusters are estimated by fitting the modified King's empirical model on their stellar density profiles. The other basic parameters of the clusters such as distance, reddening, and age are obtained by visual fitting of the Marigo's solar metallicity isochrone on their IR colour-magnitude diagrams (CMDs). The mean-proper motion of the clusters are estimated through the individual proper motion of probable members identified through the dynamical and statistical methods. The archival catalogues ($JHKW_1W_2$) are constructed for both the clusters by compiling the extracted data from the $PPMXL$ and $WISE$ catalogues. The various colour-excesses, such as $E(J-H)$, $E(H-K)$ and $E(W_1-W_2)$, are estimated using the best fit theoretical isochrone on the $(J-H)-H$, $(H-K)-H$ and $(W_1-W_2)-H$ CMDs, respectively. The ratios of various infrared colours of the clusters are obtained through their two-colour diagrams. We also identify the most probable members in these clusters by estimating spatial, kinematic and spatio-kinematic probabilities of stars within the cluster. A correlation between the $E(H-K)$ and $E(W_1-W_2)$ is also established. 
\end{abstract}


\end{frontmatter}


\def\astrobj#1{#1}
\section{Introduction}
\label{sect:intro}
The Open star clusters (OCLs) are gravitationally bound group of stars situated at the same distance from the Sun and have the similar age. They are born embedded within the giant molecular clouds and visible only at the infrared wavelengths during their initial stages of the evolution. Their embedded birthrate always exceeds to the growth-rate of the visible open star clusters \citep{Lada+2003}. The structure of the cluster is a result of its evolutionary processes such as initial physical conditions of the molecular clouds, internal structure, external tidal perturbation, etc \citep{Chen+2004, Sharma+2008}. From the analysis of cluster radius of several OCLs in optical and near-IR bands, it has been found that the cluster size is usually larger in the near-IR bands, barring few exceptions, as compared to the optical bands \citep{Sharma+2006}. Since, OCLs are embedded in the Galactic disk and are generally affected by the field star contamination, hence it is necessary to distinguish between members and non-members in the field of cluster. In recent years, the detailed membership analysis of stars in the cluster field has become a subject of intense investigation, mainly in view to understand the cluster properties \citep{Carraro+2008, Yadav+2008, Joshi+2014}.

In the present study we aim to determine the basic parameters of two OCLs, namely \astrobj{DOLIDZE~14} and \astrobj{NGC~110} using the $PPMXL$ and $WISE$ database. PPMXL \citep{Roeser+etal+2010} is a catalogue of positions, proper motions, 2MASS and optical photometry of 900 million stars and galaxies. The Two Micron All Sky Survey (2MASS;~\citet{Skrutskie+2006}) uses two highly-automated 1.3-m telescopes (one at Mt. Hopkins, Arizona (AZ), USA and other at CTIO, Chile) with a three-channel camera ($256 \times 256$ array of HgCdTe detectors in each channel). These detectors are capable of observing the sky simultaneously at J (1.25 $\mu m$), H (1.65 $\mu m$), and K (2.17 $\mu m$) bands, up to a 3$\sigma$ limiting sensitivity of 17.1, 16.4 and 15.3 $mag$ in the three bands, respectively. This survey has proven to be a powerful tool in the analysis of the structure and stellar content of open star clusters \citep{Bica+2003}. PPMXL also includes a spatially unlimited catalogue,  USNO-B1.0 \citep{Monet+2003} which lists positions, proper motions and magnitudes of stars in various pass-bands and is believed to provide all-sky coverage down to  21 $mag$ in the $V$ band. Further, the Wide-field Infrared Survey Explorer (WISE;~\citet{Wright+etal.+2010}) operating in the mid-IR is a NASA Medium Class Explorer mission which conducted a digital imaging survey of the entire sky in the 3.4, 4.6, 12 and 22 $\mu$m (referred as $W_1$, $W_2$, $W_3$ and $W_4$ passbands, respectively) and has produced a reliable Source Catalog containing accurate photometry and astrometry of more than 300 million stellar objects. The cluster $NGC~110$ is previously studied by \citet{Tadross+2011}, while $DOLIDZE~14$ has been termed as an OCL in a catalogue provided by \citet{Alter+1970} but have not been studied so far. In the present paper, we have determined various physical parameters of these OCLs in Sec. 2 and discussed the detailed membership estimation in Sec. 3. The results of physical properties of clusters and their significance have also been discussed and summarized in Sec. 4.
\section{Basic Parameters of Clusters}
The basic physical parameters of a cluster are reddening, distance, age, radius and mean-proper motion. Among them, the distance, reddening and age provide information about the cluster location from the Sun, presence of interstellar gas within cluster and its origin time from present. On the other hand, the average radial motion and extent from center are expressed on the basis of its proper-motion and the radius, respectively. Furthermore, the colour-magnitude diagrams (CMDs) and two-colour magnitude diagrams (TCDs) are the specific tools for the estimation of parameters of each cluster. For the purpose of such estimations, the finding charts of the clusters \astrobj{DOLIDZE~14} and \astrobj{NGC~110} have been constructed in Fig.~1, where the size of dots is related to the brightness of a star in $R$ passband at the $USNO-B1.0$ second epoch observation. Keeping in view the fact that the detection of all stars of a cluster region is never possible in a single photometric band because of various technical and photometric limitations, here we use the $PPMXL$ catalogue. As this catalogue is very effective to overcome the problem of incompleteness of data to a large extent since it includes the star detection in near-IR bands ($J$,$H$ and $K$) from $2MASS$ as well as magnitudes in optical bands ($B$, $R$ and $I$) from $USNO-B1.0$. With this background, the estimation procedure of various physical parameters of OCLs is described in following subsections.
\subsection{Cluster radius}
It is well known that the spatial coverage and uniformity of PPMXL photometry may be used for the determination of the cluster radius, which allows one to obtain reliable data on the projected distribution of stars for the large extensions to the cluster halos \citep{Tadross+2012}. This catalogue contains near-IR magnitude of stars (from $2MASS$) as well as photometric $BRI$ magnitude of stars (from $USNO-B1.0$) and these stars may be used to calculate the stellar density of a cluster from its center to circumference. The stellar density has been found to be maximum at the cluster center. The stellar density of clusters is obtained through the star-count method. The center of each cluster has been obtained by fitting the Gaussian function to the profile of star counts in both the right ascension ($RA$) and declination ($DEC$) directions. Thus, the center of \astrobj{NGC~110} has been found at the coordinates $(6.84352d, 71.3990d)$ or $(00^{h}:27^{m}:22.4^{s}, +71^{o}:23^{'}:56.6^{''})$ while that of \astrobj{DOLIDZE~14} has been found at the coordinates $(61.6111d, 27.3741d$ or $(04^{h}:06^{m}:26.7^{s}, +27^{o}:22^{'}:26.7^{''})$. The cluster radius is then defined as the distance from cluster center, where the cluster stellar density coincides with the field stellar density. The cluster radius is then estimated using its radial density profile (RDP), which is constructed using the stellar density  within the concentric shells located at equal incremental intervals ($r{\leq}1~arcmin$) from the cluster center. For this purpose, the density function $\rho (r)$ is given by the modified empirical King model \citep{King+1966, Kaluzny+1992}. The expression of $\rho (r)$ is given as below:
\begin{equation}
\rho (r)= f_{bg} + \frac{f_0}{1+(\frac{r}{r_c})^2}~~~,
\end{equation}
where $f_{bg}$, $f_0$, and $r_c$ are the background field stellar density, central star density and the core radius of the cluster, respectively. Here, the core radius is defined as the distance from cluster center, where the stellar density reduces to half of that at its center. The RDP of each cluster thus obtained are shown in Fig.~2. The best fit model gives the core-radius of \astrobj{NGC~110} as $0.79{\pm}0.19~arcmin$, while that of \astrobj{DOLIDZE~14} as $2.76{\pm}0.81~arcmin$. The cluster radius for \astrobj{NGC~110} is then estimated to be $5.6{\pm}0.4~arcmin$, which is about half of that estimated  by \citet{Tadross+2011}. On the other hand, the radius of \astrobj{DOLIDZE~14} cluster has been  estimated to be $9.6{\pm}0.2~arcmin$, which is a little higher as given by \citet{Alter+1970} in their catalogue. These radial extensions of the clusters under study have been depicted by circles in Fig.~1. In these estimations, the background stellar density of \astrobj{NGC~110} and \astrobj{DOLIDZE~14} have been obtained as 10.2 and 4.5 $stars/arcmin^2$, respectively.
\subsection{Field stars separation}
\label{fss}
The field stars are those foreground and background stars which are found on the direction of cluster but have different birth time and different evolution process compare to clusters stars. They are also effected the accuracy of measurement of cluster parameters. The rejection of field stars from cluster region is necessarily required to overcome their influence in estimation of cluster parameters.
\subsubsection{Dynamical approach}
As OCLs are gravitationally bound systems, the mean-proper motion values may also be used to separate the field stars from the cluster region using 3${\sigma}$ clipping\footnote{If the proper motion center of field stars and cluster stars are close to each other then the mostly field stars within cluster region have high proper motion values and lie outside 3${\sigma}$ limit}. The mean and ${\sigma}$ values of such proper motion of sky covering stars of each cluster have been determined in both the directions and are used to separate the stars not lying within the $3{\sigma}$ value in both the directions. Using iteration procedure, the stars lying outside  $3{\sigma}$ value from the mean value of proper-motion of cluster in both directions have been completely eliminated and the remaining stars (the dynamical members) are used to find the mean proper motion of sky covering stars of the clusters. For the case of \astrobj{NGC~110}, a total of 667 stars (out of 933 stars) have been used which leads to the estimation of its mean-proper motion through its dynamical members as $3.55{\pm}0.19$ $mas/yr$ and $3.06{\pm}0.17$ $mas/yr$ in RA and DEC directions, respectively. Similarly, for \astrobj{DOLIDZE~14}, a total of 932 stars (out of 1511 stars) have been used which leads to the estimation of its mean-proper motion through its dynamical members as $-0.57{\pm}0.29$ $mas/yr$ and $-8.38{\pm}0.35$ $mas/yr$ in RA and DEC directions, respectively. The vector point diagram of these clusters (in RA-DEC plane), thus obtained, has been depicted in Fig.~3, in which the dark green dots (in each panel) represent the dynamical members. 
     
\subsubsection{Statistical approach}
Though a number of field stars get separated from cluster region through the dynamical approach, however, some of the field stars with their proper motion lying very close to the mean proper motion of the cluster members still need to be separated from the cluster region. It is to be further emphasised that the IR magnitude of all of the dynamical members are not available in the $2MASS$ and $WISE$ archive databases due to the incompleteness of the photometry. Such, field star separation is then carried out by a statistical approach which is based on the comparison of field region stars with stars of cluster region on the CMD. For this purpose, for each cluster, we start by considering a field area as a region with its area equivalent to that of the cluster region and compare the stars occurring in these two equivalent areas using a grid around every field star in the associated colour magnitude diagrams. Such comparison, in fact, strictly depends on the size of the grid. Since, in the present approach, most of the field stars have already been removed through the dynamical approach, we prefer to choose the grid size as minimum as possible so as to reduce the possibility of rejection for the genuine cluster members. Therefore, we have removed the cluster region stars lying within a grid of colour-magnitude space given by $(H,J-H)=({\pm}0.05,{\pm}0.01)$ corresponding to each field star. We have thus found 321 stars in statistically cleaned $(J-H)-H$ CMD of \astrobj{DOLIDZE~14} and 234 stars in that of \astrobj{NGC~110}.

\subsection{Mean proper-motion}
The mean proper motion of a cluster is defined as the average angular speed of cluster per year by which it has changed its position over the sky. For this purpose, the PPMXL catalogue \citep{Roeser+etal+2010} provides the proper-motion values for various stars in both $RA$ and $DEC$ directions. Employing the two field stars subtraction methods described above, we yield a total of 226 and 302 probable members of the clusters \astrobj{NGC~110} and \astrobj{DOLIDZE~14}, respectively. The stars have then been used for the estimation of the mean proper motion of individual clusters. We have thus found the mean-proper motion of \astrobj{NGC~110} as $4.03{\pm}0.29~mas/yr$ and $2.53{\pm}0.23~mas/yr$ in $RA$ and $DEC$ directions, respectively while the same for cluster \astrobj{DOLIDZE~14} are found to be $-0.15{\pm}0.34~mas/yr$ and $-7.79{\pm}0.41~mas/yr$. The probable members for estimating the mean proper motion of clusters are depicted by blue dots in Fig.~3.
\subsection{Two-colour diagrams (TCD) and colour-excess}
The two-colour diagrams (TCD) are very useful to estimate the relation of various colour-excesses and their variations towards the direction of cluster region. In the present study, these diagrams have been constructed using the stars with the estimation-error of magnitude less than 0.1 mag in IR bands. As a result, we have found 414 and 456 stars of \astrobj{NGC~110} and 461 and 952 stars of \astrobj{DOLIDZE~14} in the $2MASS$ and $WISE$ catalogues, respectively. Among these two catalogues, the number of common stars in \astrobj{NGC~110} are found to be 252 while those in \astrobj{DOLIDZE~14} are 384. Among these common stars, the number of probable members are estimated to be 136 and 254 for \astrobj{NGC~110} and \astrobj{DOLIDZE~14}, respectively. Such members are depicted by red dots in each TCD. Using $2MASS$ and $WISE$ archival data, the different types of TCDs such as  $(H-J)-(H-K)$, $(H-W_1)-(H-K)$ and $(H-W_2)-(H-K)$ for each individual cluster have been plotted in Fig.~4.  Similarly, $(J-H)-(W_1-W_2)$ plots of each cluster have also been depicted in Fig.~5. The blue dots in each TCD represent the members of cluster with their spatio-kinematic probability as more than 0.5 and the magnitude estimation error as less than 0.1 mag (Most Probable members i.e. MPMs). The blue dashed lines in each panel of Figs.~4 and 5 show the best linear fit between two different colour-excesses. The slope of different TCDs are listed in Table~1. The $(H-J)-(H-K)$ diagram of each cluster then provides the ratio, $\frac{E(J-H)}{E(H-K)}$, which is found to be approximately equal to $1.49{\pm}0.33$ and $1.40{\pm}0.19$ for \astrobj{NGC~110} and \astrobj{DOLIDZE~14}, respectively. These values are in close agreement with $\frac{E(J-H)}{E(H-K)}=\frac{5}{3}$ obtained through the Mechure relation \citep{Mechure+1970} within their respective errors. The reddening, $E(B-V)$, is estimated using the following relations \citep{Fiorucci+2003}:
$$E(J-H) = 0.31~E(B-V)$$     
$$E(J-K) = 0.48~E(B-V)$$
The different colour-excesses, $E(J-H)$, $E(H-K)$ and $E(W_1-W_2)$ for each cluster are obtained through $(J-H)-H$, $(H-K)-H$ and $(W_1-W_2)-H$ CMDs, as shown in Fig.~6.  The estimated reddening value is found to be $0.42{\pm}0.03$ mag for  \astrobj{NGC~110} which is close to $0.46 {\pm} 0.10$ mag as estimated by \citet{Tadross+2011} while the same for  \astrobj{DOLIDZE~14} has been found to be $0.32{\pm}0.02$ mag.
\subsection{Colour-Magnitude Diagrams: Distance, age and colour-excess}
The identification of Main Sequence (MS) in CMDs allows us to determine the model-dependent mass, radius and distance of each cluster \citep{Joshi+2014}. For the estimation of the distance and age of different clusters, we use different type of CMDs such as $(J-H)-H$, $(H-K)-H$ and $(W_2-W_1)-H$ as shown  in Fig 6. The black dots in Fig 6 show the stars obtained after separating the field stars from the cluster region. The parameters of each cluster are then obtained through fitting of Marigo isochrones \citep{Marigo+2008} of solar metallacity on different CMDs  of clusters. The best fit isochrone then gives the value of apparent distance-modulus as $10.8{0.3}$ mag and $11.3{0.2}$ mag for \astrobj{NGC~110} and \astrobj{DOLIDZE~14}, respectively. The reddening correction has been obtained by the absorption formula $A_H /A_V = 0.176$ as suggested by \cite{Dutra+2002}. The distance of each cluster is then determined through the following relation:
\begin{equation}
D=10^{[m_{0}-R_{H} ~E(J-H)+5]/5}
\end{equation}
where $R_{H}$ $\sim$ 1.77 and $m_{0}$ is visual distance modulus. Here, $m_{0}-R~ E(J-H)$ is also identified as the distance-modulus ($m_{H}$). We have estimated the value of distance modulus for \astrobj{DOLIDZE~14} as $11.12{\pm}0.18$ mag while it turns out to be $10.57{\pm}0.28$ mag for \astrobj{NGC~110}. The distance is then obtained as $1.29{\pm}0.22~kpc$ for \astrobj{NGC~110} which is close to $1150{\pm}53~pc$ as calculated by \cite{Tadross+2011} while the same for the \astrobj{DOLIDZE~14} comes out to be $1.67{\pm}0.14~kpc$. In addition, the best fit of Marigo isochrones further leads to the log(age) of \astrobj{NGC~110}  as 9.0 (equivalent to $1~Gyr$) which is close to the value $0.9{\pm}0.04~Gyr$ as estimated by \citet{Tadross+2011}. The log(age) of \astrobj{DOLIDZE~14} has been estimated as 9.1 which is equivalent to $1.26{\pm}0.08~Gyr$. Furthermore, we have also estimated the different colour-excesses such as, $E(H-K)=0.08$ and $E(W_1-W_2)=0.01$, by best fitted theoretical isochrones on $(H-K)-H$ and $(W_2-W_1)-H$ CMDs of \astrobj{DOLIDZE~14}. The similar estimations for \astrobj{NGC~110} have been found to be, $E(H-K)=0.06$ and $E(W_1-W_2)=0.01$. As a result, these two colour-excess for the presented studied clusters are corelated as,
\begin{equation}
\frac{E(H-K)}{E(W_1-W_2)} \simeq 6
\end{equation} 
In addition, it has also been found that the ratio $E(J-H)/E(H-K)$ obtained for these clusters using best fitted isochrones on $(J-H)-H$ and $(H-K)-H$ CMDs agrees well with the value obtained through the \cite{Mechure+1970} relations.

\section{Membership probabilities} 
It is a well known fact that the cluster region is influenced by the foreground and background stars. It is, therefore, required to define a factor which may indicate the possibility of a star belonging to a particular cluster region. It is defined as the membership probability of the star, which may be calculated by the star position from the cluster center and the proper motion of a star related to mean proper motion value of the cluster. The probability associated with the position is defined as the spatial probability, while that related to the proper motion is termed as the kinematic probability. The criteria based on these probabilities is expected to yield the most probable members of a cluster. The location of these most probable members in cluster CMDs then becomes a tracing tool of the best visual fitted isochrones. As such, the probabilities needed for the estimation of membership of stars within a clusters, may be described as follows.

The probability of localizing a star in the cluster field, is termed as the spatial probability, which may be given in the following form \citep{Joshi+2012}:
 \begin{equation}
 p_{sp} = 1 - \frac{r_i}{r_c}
 \end{equation}
where $r_i$ is the angular distance of $i^{th}$ star from the center of the cluster and $r_c$ is the cluster radius. For the stars lying outside the cluster region, such probability always has the negative value.

It is well known that the proper motion of member stars of a cluster usually differs the values from the mean-proper motion of the cluster. As a result, the  deviation of proper-motion of a member in both $RA$ and $DEC$ directions with respect to the mean-proper motion of cluster provides a measure of its kinematic probability. The kinematic probability of stars as has been derived using the technique given by \citet{Kharchenko+2004}, may be given by, 
\begin{equation}
p_k=exp{\left[-0.25\{(\mu_{x}-\bar{\mu_{x}})^2/\sigma_{x}^2 + (\mu_{y}-\bar{\mu_{y}})^2/\sigma_{y}^2\}\right]}
\end{equation} 
where $\sigma_{x}^2=\sigma_{\mu x}^2+\sigma_{\bar{\mu x}}^2$ and $\sigma_{y}^2=\sigma_{\mu y}^2+\sigma_{\bar{\mu y}}^2$. The $\mu_{x}$ and $\mu_{y}$ are the proper motion of any star in the direction of $RA$ and $DEC$, respectively, while $\sigma_{\mu x}$ and $\sigma_{\mu y}$ are the uncertainties in measurement of the proper motion of stars along the respective directions.

The two type of probabilities discussed above for the determination of the membership of a star in a cluster may sometime lead to conflicting results. It may be due to the fact that the spatial probability, in fact, drops continuously from the center to cluster circumference while the kinematic probability grows independent of the spatial configuration of the stars in a cluster field. As a result, one has to heavily constrain the cluster system to match the most probable membership results. Moreover, a cluster system is also expected to be a dynamical system with varying proper motion of its constituent member stars. Hence, in order to avoid any inconsistency in membership determination from the associated probability, it is naturally desired to redefine the probability in such a way that it includes the spatial feature of the members on one hand and the dynamical evolution on the other. It may be identified as the ``spatio-kinematic probability" which may be expressed by a geometrical mean as given below;
\begin{equation}
P_{sk}=(p_{sp} \times p_k)^{\frac{1}{2}} \\
= [(1 - \frac{r_i}{r_c})(e^{-0.25\{(\mu_{x}-\bar{\mu_{x}})^2/\sigma_{x}^2 + (\mu_{y}-\bar{\mu_{y}})^2/\sigma_{y}^2\}})]^{\frac{1}{2}}
\end{equation}\\
Using the above expression, we identified the stars belonging to the clusters having $P_{sk}$ more than 0.5. Out of this, we selected only those stars as a most probable members whose magnitude estimation error is less than 0.1 mag in each corresponding band. We give the magnitude and probability of most probable members identified within the clusters \astrobj{NGC~110} and \astrobj{DOLIDZE~14} in Tables~2 and 3, respectively. Such most probable members have shown to closely follow the main sequence of cluster in the CMD and are depicted by big blue dots in Fig.~7.
\subsection{Determination of effectiveness of probabilities}
It is well known that the stellar density of cluster field is affected by the presence of foreground and background stars. We may, therefore, also calculate the effectiveness of membership determination using the expression given below \citep{shao+1996}:
\begin{equation}
E=1-\frac{N \sum_{i=1}\limits^{N} {[ P_{i}(1-P_{i}) ]} }{{\sum_{i=1}\limits^{N}}P_{i}{\sum_{i=1}\limits^{N}}(1-P_{i})}~~~,
\end{equation}
where $N$ is the total number of stars under consideration for the determination of the membership probability and $P_{i}$ indicates the probability of $i^{th}$ star of the cluster. We have found the effectiveness ($E$) for \astrobj{NGC~110} and \astrobj{DOLIDZE~14} as $0.12$ and $0.20$, respectively. The result thus fall well within the lower limit for E (0.20) as that estimated by  \citet{shao+1996} for a sample of 43 OCLs although they modified the lower limit (of $E$) to a value 0.12 for the smaller cluster.
\section{Discussion and Conclusions}
\label{sect:Conclusion}
In view of the potential importance of the open star cluster and their catalogues in understanding the evolution of the Galaxy, the present work deals with the estimation of the basic physical parameters of two clusters, viz. \astrobj{NGC~110} and \astrobj{DOLIDZE~14} using  $PPMXL$ catalogue. In the present study, the center coordinates of \astrobj{NGC~110} and \astrobj{DOLIDZE~14} are found as $(00^{h}:27^{m}:22.4^{s}, +71^{o}:23^{'}:56.6^{''})$ and $(04^{h}:06^{m}:26.7^{s}, +27^{o}:22^{'}:26.7^{''})$, respectively. The cluster radius of  \astrobj{NGC~110} and \astrobj{DOLIDZE~14} estimated using the King empirical model, and are found to be 5.2$\pm$0.4 $arcmin$  and 9.6$\pm$0.2 $arcmin$, respectively. The radial density profiles of these clusters indicate that these clusters are not densely populated. The mean-proper motion of \astrobj{NGC~110} and \astrobj{DOLIDZE~14} in their $RA$ and $DEC$ directions were estimated as ($4.03{\pm}0.29$ and $2.53{\pm}0.23$ $mas/yr$) and ($-0.15{\pm}0.34$ and $-7.79{\pm}0.41$ $mas/yr$), respectively. The reddening, $E(B-V)$, is estimated as $0.42{\pm}0.03$ $mag$ for \astrobj{NGC~110} and $0.32{\pm}0.02$ $mag$ for \astrobj{DOLIDZE~14}. However, $E(B-V)$ estimated using the Fiorucci \& Munari (2003) relation is still needed to be confirmed by further investigation through $UBVRI$ $CCD$ photometry. The best fit theoretical isochrone on $JHK$ catalogue of \astrobj{NGC~110} yielded a distance of $1.29{\pm}0.22~kpc$ for the cluster, while it is estimated as $1.67{\pm}0.14~kpc$ for \astrobj{DOLIDZE~14}. In addition, the log(age) estimated for these clusters (approx. 9) demonstrates that these clusters belong to old-age categories of OCLs. The Table~4 summerises all the estimated parameters of the present studied clusters. For the identification of most probable members of the clusters, the statistical approach of field star separation has been shown to further limit the probable members and such most probable members of these clusters are considered to be the true tracer of theoretical isochrones.

The colour-excesses $E(H-K)$ and $E(W_1-W_2)$ obtained by best fitted isochrones on various CMDs of the cluster under study have also been found to be linearly correlated as, given by eq.(3). Similar correlations of these clusters in other bands may also be drawn by using $UBVRI$ data. A deep $UBVRI$ $CCD$ photometric study of these clusters is further needed to investigate some of their important dynamical properties, viz, mass-segregation, luminosity-function, etc.
\section*{Acknowledgments} 
GCJ is thankful to University Grants Commission (New Delhi) for the financial assistance in terms of a RFSMS fellowship. This publication makes use of data products from the Two Micron All Sky Survey, which is a joint project of the University of Massachusetts and the Infrared Processing and Analysis Center/California Institute of Technology, funded by the National Aeronautics and Space Administration and the National Science Foundation. This publication makes use of data products from the Wide-field Infrared Survey Explorer, which is a joint project of the University of California, Los Angeles, and the Jet Propulsion Laboratory/California Institute of Technology, funded by the National Aeronautics and Space Administration.
\bibliographystyle{model1a-num-names}
\bibliography{<your-bib-database>}

\begin{thebibliography}{}


\bibitem[Alter et~al.(1970)Alter et~al.]{Alter+1970}
Alter, G., Balazs, B. $\&$ Ruprecht, J., 1970, Catalogue of Star Clusters and associations, $2^{nd}$ Ediation, Akademial Kiada Budapest, printed in Hungary(card form)

\bibitem[Bica \& Bonatto(2003)Bica $\&$ Bonatto]{Bica+2003}
Bica, E., Bonatto, Ch., Dutra, C.~M., 2003, A{\&}A, 405, 991

\bibitem[Carraro et~al.(2008)Carraro et~al.]{Carraro+2008}
Carraro, G., Villanova S., Demarque P., Moni-Bidin C., McSwain M. ~V., 2008, MNRAS, 386, 1625

\bibitem[Chen et al.(2004)Chen et~al.]{Chen+2004}
Chen, W.~P., Chen, C.~W., Shu, C.~G., 2004, AJ, 128, 2306

\bibitem[Dutra et al.(2002)Dutra et al.]{Dutra+2002}
Dutra, C.M., Santiago, B.X., Bica, E., 2002, A{\&}A, 383, 219

\bibitem[Fiorucci \& Munari(2003)Fiorucci \& Munari]{Fiorucci+2003}
Fiorucci, M., Munari, U., 2003, A$\&$A, 401, 781

\bibitem[Joshi et~al.(2012)Joshi et~al.]{Joshi+2012}
Joshi, Y.~C., Joshi, S., Kumar, Brijesh, Mondal, Soumen, Balona, L.~A., 2012, MNRAS, 419, 2379

\bibitem[Joshi et~al.(2014)Joshi et~al.]{Joshi+2014}
Joshi, Y.~C., Balona, L.~A., Joshi, S., Kumar, B., 2014, MNRAS, 437, 804

\bibitem[Lada \& Lada(2003)Lada \& Lada]{Lada+2003}
Lada, C.~J., Lada, E.~A., 2003, ARA$\&$A, 41, 57

\bibitem[King(1966)King]{King+1966}
King, I., 1966, AJ, 71, 64

\bibitem[Kaluzny $\&$ Udalski(1992)Kaluzny $\&$ Udalski]{Kaluzny+1992}
Kaluzny, J., Udalski A., 1992, Acta Astro., 42, 29

\bibitem[Kharchenko et~al.(2004)Kharchenko et~al.]{Kharchenko+2004}
Kharchenko N. ~V., Piskunov A. ~E., Roeser S., Schilbach E., Schola R.~D., 2004, Astron. Nachr., 325, 740

\bibitem[Marigo et~al.(2008)Marigo et~al.]{Marigo+2008}
Marigo P., Girdardi L., Bressan A., et al., 2008, A$\&$A, 482, 883

\bibitem[Mechure(1970)Mechure]{Mechure+1970}
Mechure, R.~D., 1970, AJ, 75, 41

\bibitem[Monet et~al.(2003)Monet et~al.]{Monet+2003}
Monet, D. G., Levine, S. E., Canzian, B., al., 2003, AJ, 125, 984

\bibitem[Roeser et~al.(2010)Roeser et~al.]{Roeser+etal+2010}
Roeser, S., Demleitner, M., Schilbach, E., 2010, AJ, 139, 2440

\bibitem[Shao $\&$ Zhao(1996)Shao $\&$ Zhao]{shao+1996}
Shao, Z.~Y., Zhao, J.~L., 1996, Acta Astron. Sin., 37, 377

\bibitem[Sharma et~al.(2006)Sharma et~al.]{Sharma+2006}
Sharma, S., Pandey, A. K., Ogura, K., Mito, H., Tarusawa, K., Sagar, R., 2006, AJ, 132, 1669

\bibitem[Sharma et~al.(2008)Sharma et~al.]{Sharma+2008}
Sharma, S., Pandey, A.~K., Ogura, K., Taori, A., Pandey, K., Sandhu, T.~S., Sagar, R., 2008, AJ, 135, 1934

\bibitem[Skrutskie et~al.(2006)Skrutskie et~al.]{Skrutskie+2006}
Skrutskie, M.~F., Cutri, R.~M., Stiening, R., et al., 2006, AJ, 131, 1163

\bibitem[Tadross(2011)Tadross]{Tadross+2011}
Tadross, A.~L., 2011, JKAS, 44, 1

\bibitem[Tadross(2012)Tadross]{Tadross+2012}
Tadross, A.~L., 2012, New Astron., 17, 198

\bibitem[Wright et~al.(2010)Wright et~al.]{Wright+etal.+2010}
Wright, E.~L., Eisenhardt, P.~R.~M., Mainzer, A.~K., et al. 2010, AJ, 140, 1868

\bibitem[Yadav et~al.(2008)Yadav et~al.]{Yadav+2008}
Yadav R.~K.~S., Bedin L.~R., Piotto G., et al., 2008, A{\&}A, 484, 609

\end{thebibliography}

\begin{figure}
   \centerline{\includegraphics[width=14.0cm]{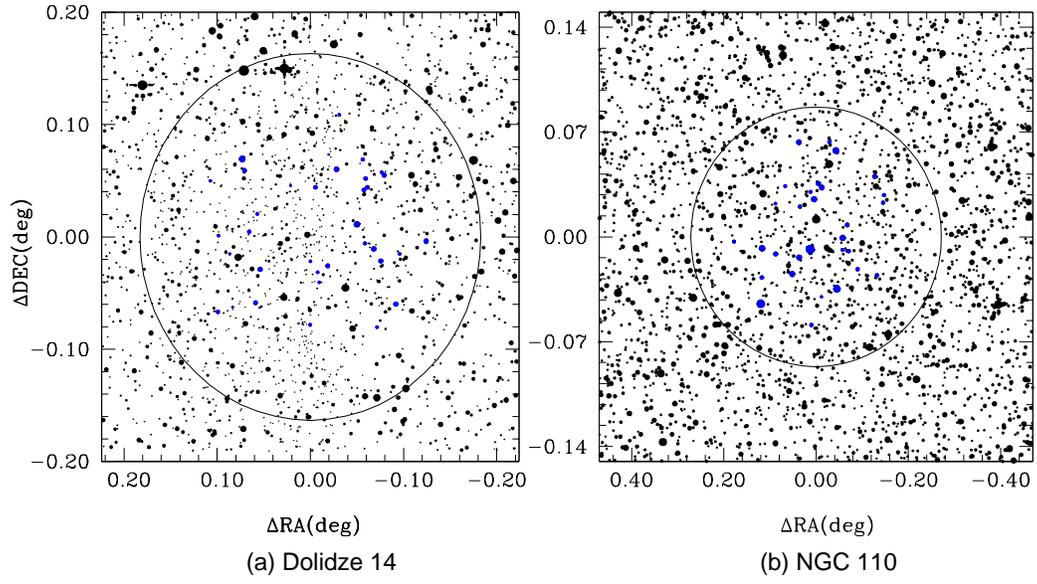}}
   \caption{ The finding charts of stars in, (a) \astrobj{DOLIDZE~14}, and (b) \astrobj{NGC~110}. The biggest size dots represent $10^{th}$ magnitude star and smallest size dots represent stars fainter than 18 mag. The blue dots are identified as most probable members (MPM's) within the cluster region.} 
   \label{Fig1}
   \end{figure}
\begin{figure}
   \centerline{\includegraphics[height=14.0cm]{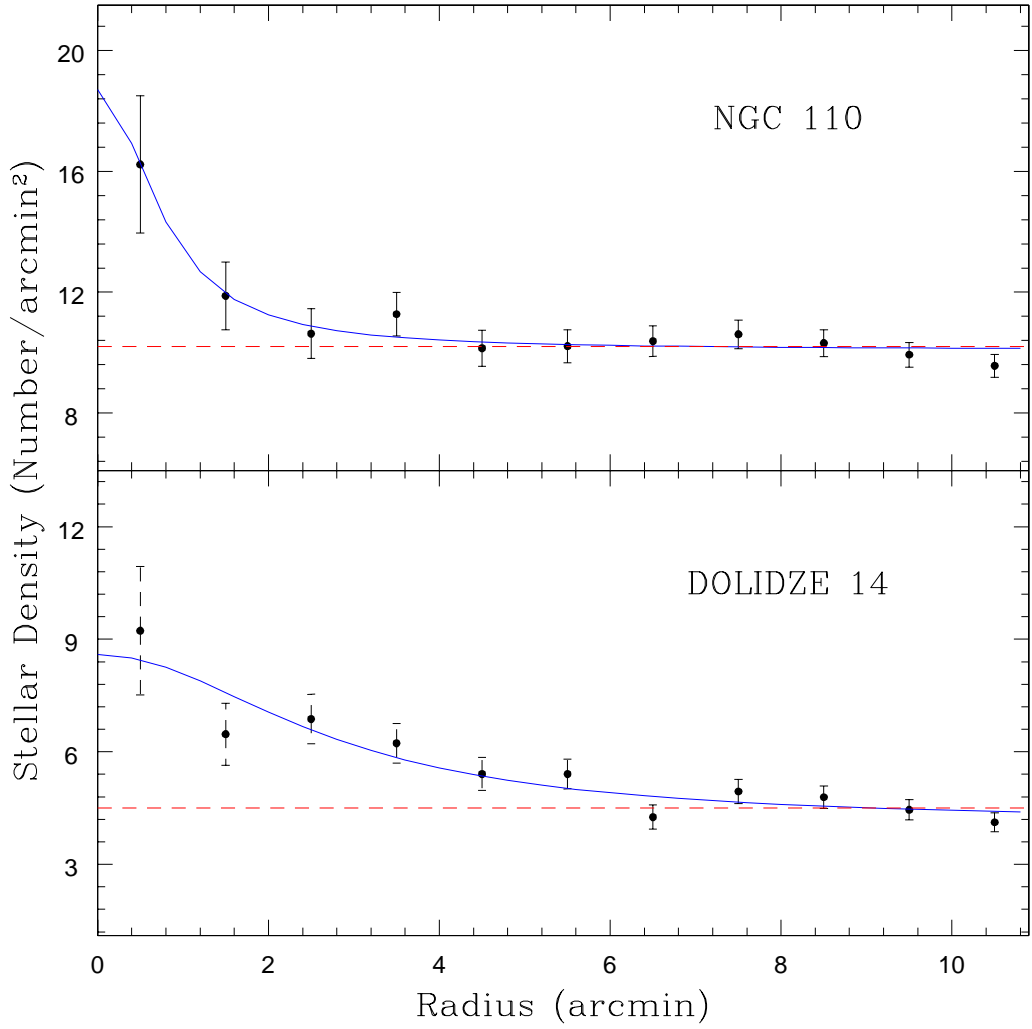}}
   \caption{ The radial density profile for the cluster \astrobj{NGC~110} and \astrobj{DOLIDZE~14}. The red dashed lines mark the background field star density including 1-$\sigma$ error. The blue solid lines mark the King model fit for radial density profiles.} 
   \label{Fig2}
   \end{figure}
 \begin{figure}
\centerline{\includegraphics[width=14.0cm]{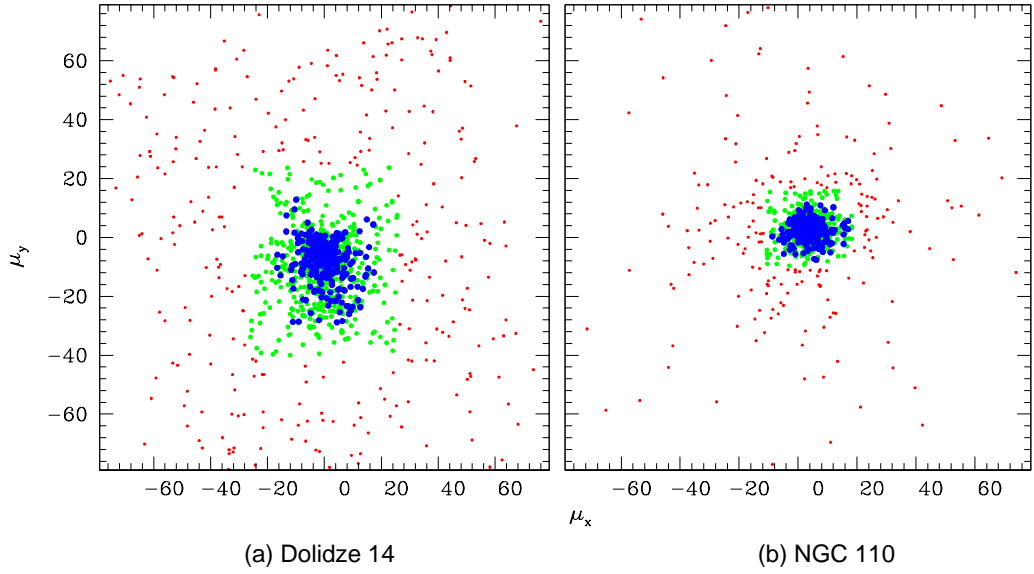}}
   \caption{ The proper-motion distribution of stars for \astrobj{NGC~110} and \astrobj{DOLIDZE~14}. The green dots represent the dynamical members used to determine the mean-proper motion while the blue  dots represent the probable members used to determine the mean proper motion of the cluster.} 
   \label{Fig3}
   \end{figure} 
 \begin{figure}
   \centerline{\includegraphics[width=14.0cm]{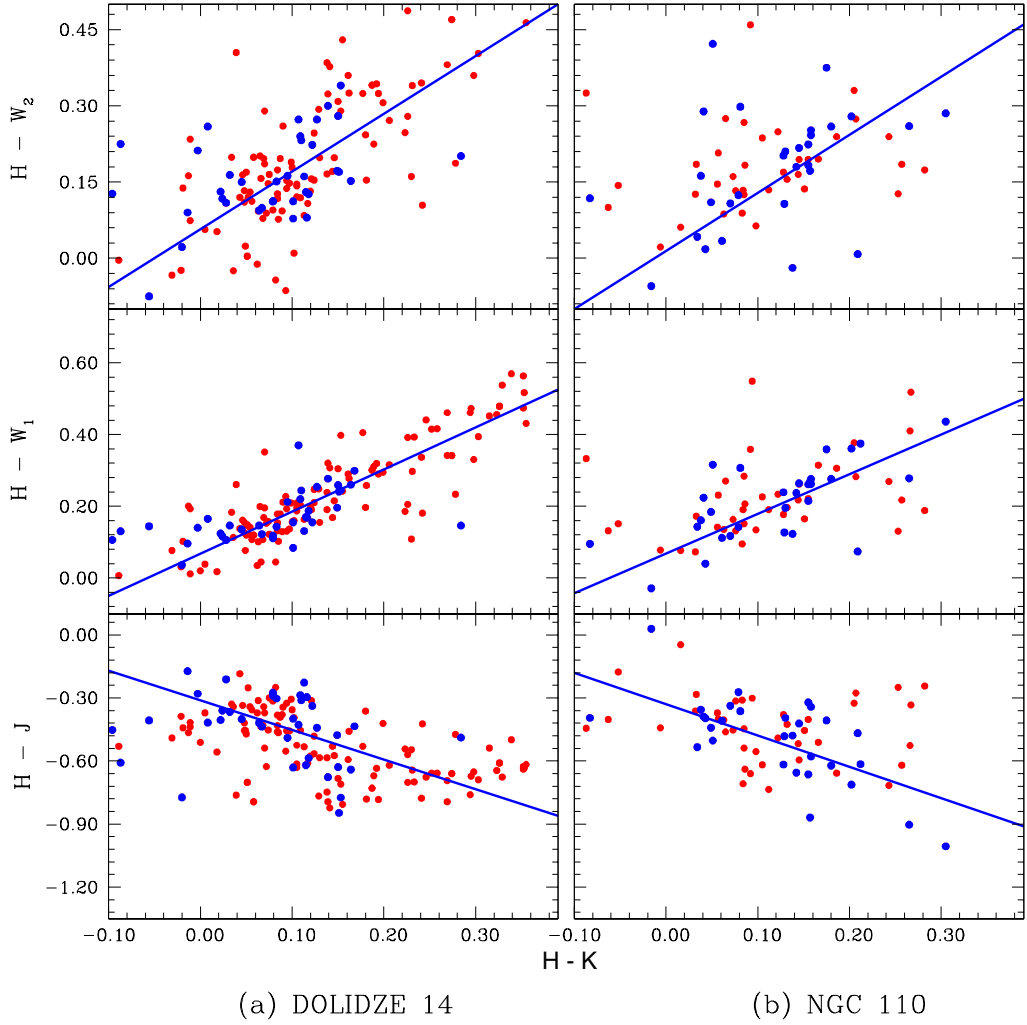}}
   \caption{ The left and right panels represent $(H-{\Lambda})$ vs $(H-K)$ TCDs for $\astrobj{NGC~110}$ and $\astrobj{DOLIDZE~14}$ (${\Lambda}$ being either $J$, $W_1$ or $W_2$). The blue solid line represents linear fit in each panel while blue dots represent the most probable members ($p_{sk}$ $>$ 0.5)} 
   \label{Fig4}
   \end{figure}
\begin{figure}
   \centerline{\includegraphics[width=14.0cm]{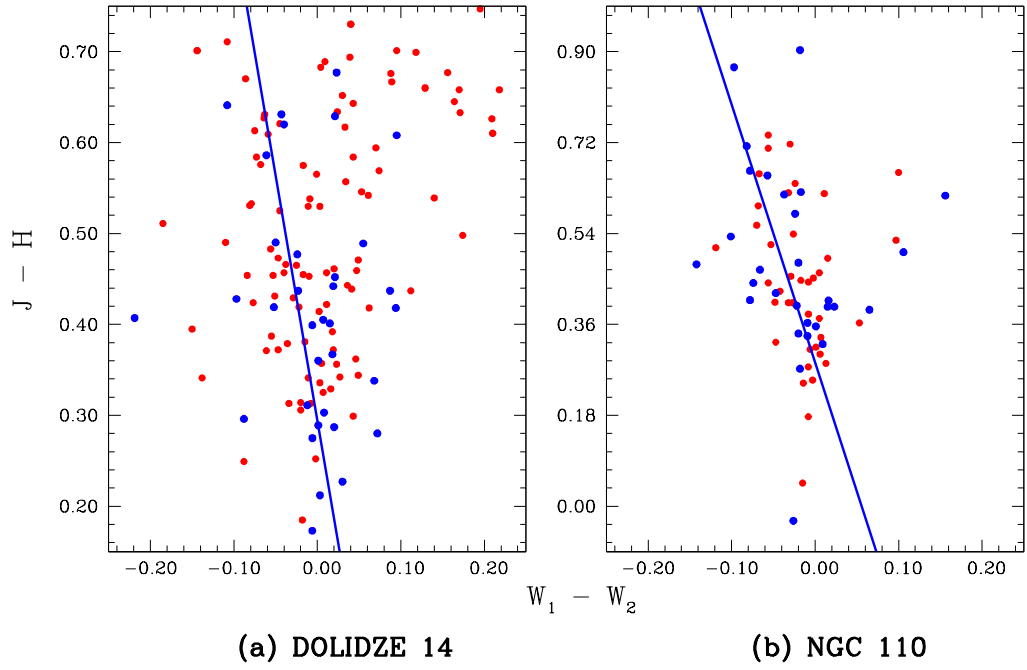}}
   \caption{ The left and right panels represent $(W_1-W_2)$ vs $(J-H)$ TCDs for \astrobj{NGC~110} and \astrobj{DOLIDZE~14} respectively. The blue solid line represents linear fit in each panel while blue dots represent the most probable members ($p_{sk}$ $>$ 0.5)} 
   \label{Fig5}
   \end{figure}  
\begin{figure}
   \centerline{\includegraphics[width=14cm]{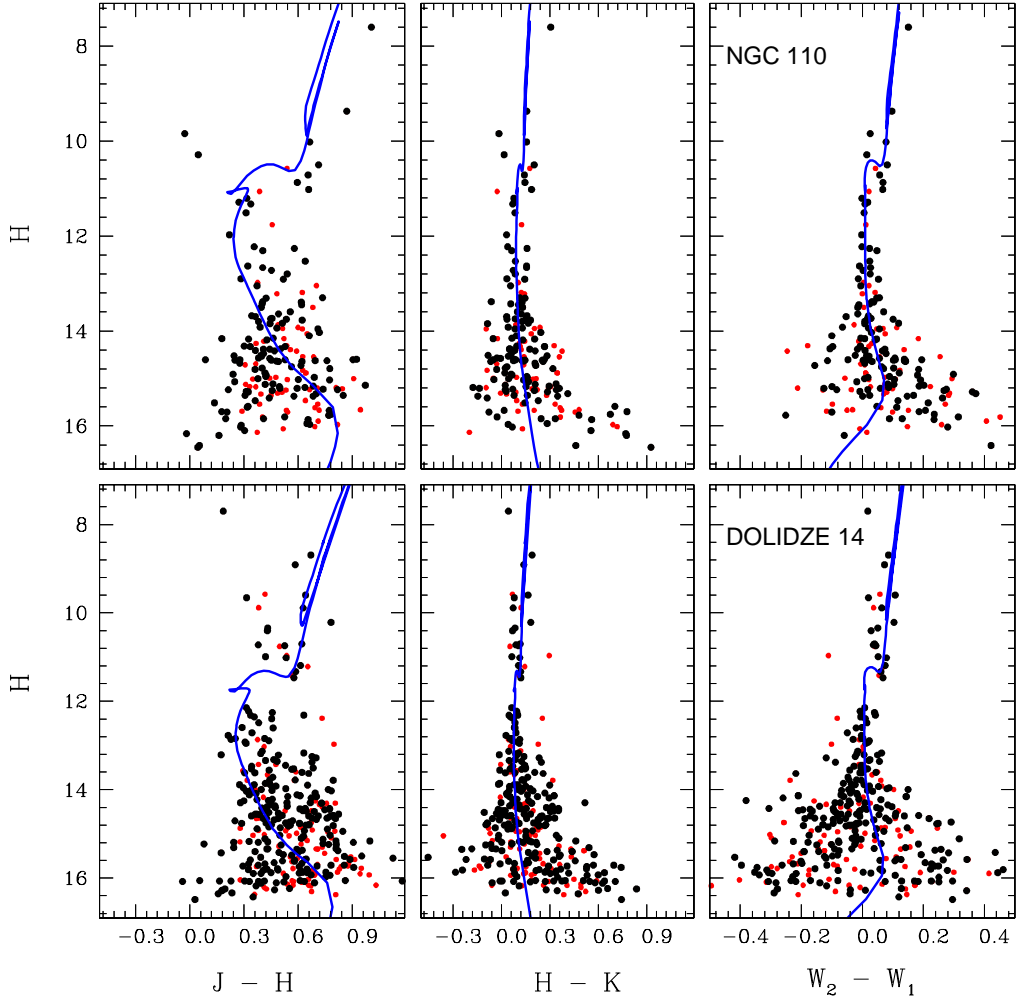}}
   \caption{ The upper and bottom panels represent colour-magnitude diagrams for \astrobj{NGC~110} and \astrobj{DOLIDZE~14} respectively. The different colours such as $(J-H)$, $(H-K)$ and $(W_1-W_2)$ are shown as a function of $H$ magnitude of stars in the left, middle and right panels respectively. The black dots on each panel represent the stars in cluster region after field stars subtraction through statistical approach while red dots represent the dynamical members of the cluster.} 
   \label{Fig6}
   \end{figure}
\begin{figure}
   \centerline{\includegraphics[width=14cm]{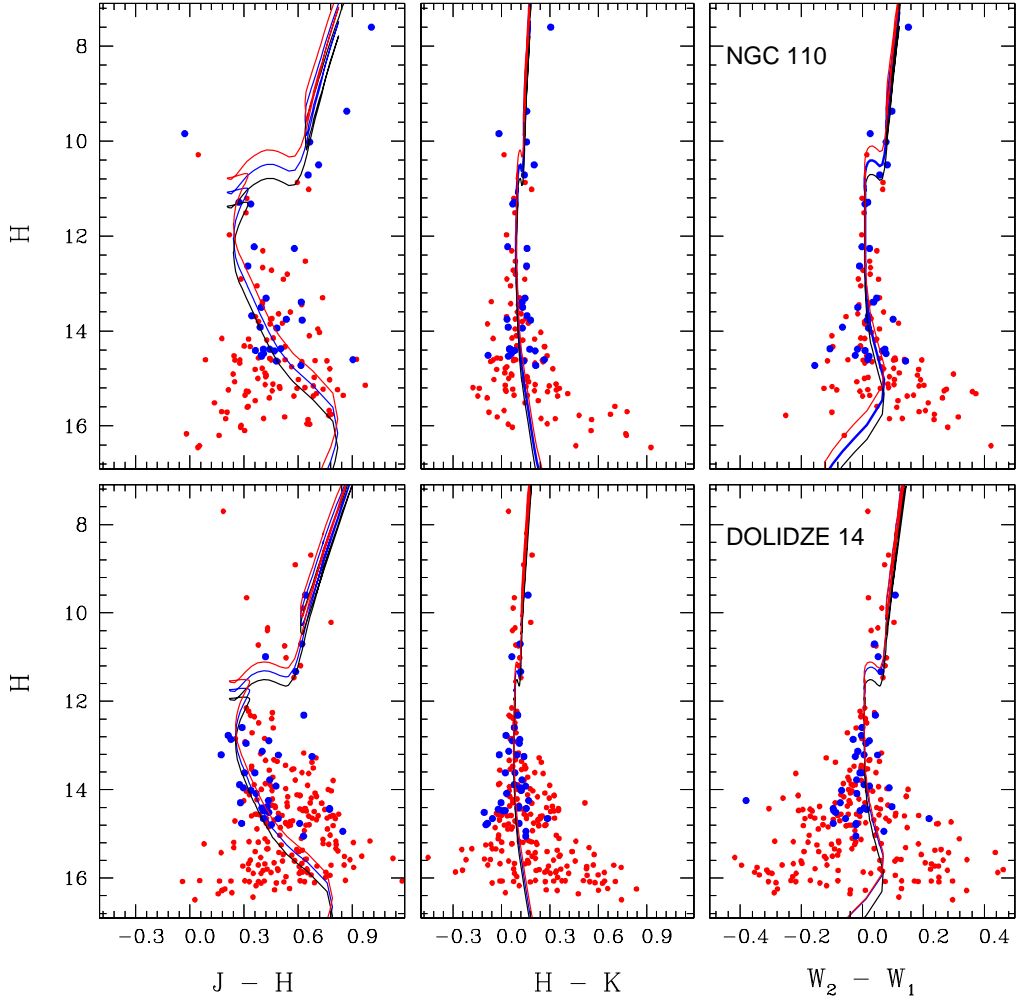}}
   \caption{ The upper and bottom panels represent CMDs for $\astrobj{NGC~110}$ and $\astrobj{DOLIDZE~14}$, respectively. The different colours such as $(J-H)$, $(H-K)$ and $(W_1-W_2)$ are shown as a function of $H$ magnitude of stars in left, middle and right panels, respectively. The blue dots on each panel represent the stars having less than 0.1 mag photometric error in each band and spatio-kinematic probability greater than 0.5 for \astrobj{NGC~110} and $DOLIDZE~14$ while red dots represent the probable members of the cluster.} 
   \label{Fig7}
   \end{figure}
\newpage
\newpage
 \begin{table}
\small
\caption{Colour-ratios for the clusters under study in different IR-bands. \label{tab1}}
    \begin{center}
      \begin{tabular}{ccc} \hline
Colour Ratio& \astrobj{NGC~110} & \astrobj{DOLIDZE~14}\\
     \hline
     \\
$\frac{(J-H)}{(H-K)}$  & $1.49{\pm}0.33$ & $1.40{\pm}0.19$ \\
\\
$\frac{(H-W_1)}{(H-K)}$& $1.11{\pm}0.14$ & $1.18{\pm}0.07$ \\
\\
$\frac{(H-W_2)}{(H-K)}$& $1.13{\pm}0.53$ & $1.15{\pm}0.29$ \\
\\
$\frac{(J-H)}{(W_1-W_2)}$& $-5.09{\pm}1.16$ & $-5.43{\pm}1.65$ \\
\\
\hline
\end{tabular}
\end{center}
\end{table}
\newpage
\begin{table}
\tiny
\caption{The $JHKW_1W_2$ catalogue of stars  in \astrobj{NGC~110}, having spatio-kinematic probability greater than 0.5. \label{tab2}}
\setlength{\tabcolsep}{2.5pt}
\tiny
\medskip
\begin{center}
 \begin{tabular}{cccccccccccccccccccccccc}
  \hline
RA&& DEC&& $J$ $\pm$ ${\sigma}_J$ &&$H$ $\pm$ ${\sigma}_H$ && $K$ $\pm$ ${\sigma}_K$ && $W_1$ $\pm$ ${\sigma}_{W_1}$ && $W_2$ $\pm$ ${\sigma}_{W_2}$ &&  $p_{sk}$\\
  \hline
6.963359  && 71.354309 && 8.605${\pm}$0.048 && 7.599${\pm}$0.053 && 7.294${\pm}$0.023 && 7.163${\pm}$0.029 && 7.314${\pm}$0.021 && 0.55\\
6.798335  && 71.364241 && 10.242${\pm}$0.022 && 9.373${\pm}$0.026 && 9.216${\pm}$0.022 && 9.104${\pm}$0.024 && 9.201${\pm}$0.022 && 0.77\\
6.855886  && 71.390639 && 9.810${\pm}$0.022 && 9.839${\pm}$0.029 && 9.855${\pm}$0.022 && 9.868${\pm}$0.022 && 9.894${\pm}$0.021 && 0.61\\
6.800651  && 71.456568 && 10.686${\pm}$0.022 && 10.022${\pm}$0.028 && 9.867${\pm}$0.022 && 9.761${\pm}$0.023 && 9.839${\pm}$0.021 && 0.60\\
6.847568  && 71.424183 && 11.213${\pm}$0.024 && 10.500${\pm}$0.026 && 10.298${\pm}$0.020 && 10.139${\pm}$0.022 && 10.221${\pm}$0.020 && 0.74\\
6.895260  && 71.374188 && 11.368${\pm}$0.022 && 10.713${\pm}$0.029 && 10.571${\pm}$0.022 && 10.476${\pm}$0.024 && 10.533${\pm}$0.021 && 0.61\\
6.785487  && 71.398294 && 11.562${\pm}$0.024 && 11.290${\pm}$0.029 && 11.211${\pm}$0.022 && 11.148${\pm}$0.026 && 11.166${\pm}$0.023 && 0.74\\
6.960526  && 71.391339 && 11.661${\pm}$0.022 && 11.324${\pm}$0.026 && 11.254${\pm}$0.023 && 11.207${\pm}$0.024 && 11.216${\pm}$0.020 && 0.61\\
6.880143  && 71.385233 && 12.584${\pm}$0.026 && 12.228${\pm}$0.031 && 12.190${\pm}$0.026 && 12.067${\pm}$0.026 && 12.066${\pm}$0.024 && 0.51\\
6.880807  && 71.462176 && 12.841${\pm}$0.026 && 12.262${\pm}$0.032 && 12.104${\pm}$0.028 && 11.986${\pm}$0.025 && 12.010${\pm}$0.023 && 0.51\\
6.830988  && 71.431911 && 12.955${\pm}$0.026 && 12.634${\pm}$0.032 && 12.479${\pm}$0.025 && 12.419${\pm}$0.028 && 12.410${\pm}$0.028 && 0.68\\
6.838362  && 71.434640 && 13.727${\pm}$0.031 && 13.305${\pm}$0.039 && 13.160${\pm}$0.038 && 13.041${\pm}$0.030 && 13.088${\pm}$0.033 && 0.76\\
6.772891  && 71.389527 && 14.006${\pm}$0.033 && 13.389${\pm}$0.038 && 13.261${\pm}$0.040 && 13.150${\pm}$0.029 && 13.187${\pm}$0.032 && 0.81\\
6.931092  && 71.387663 && 13.897${\pm}$0.027 && 13.502${\pm}$0.041 && 13.372${\pm}$0.040 && 13.307${\pm}$0.029 && 13.292${\pm}$0.032 && 0.80\\
6.716714  && 71.439326 && 14.021${\pm}$0.028 && 13.679${\pm}$0.044 && 13.521${\pm}$0.049 && 13.417${\pm}$0.026 && 13.437${\pm}$0.033 && 0.55\\
6.878371  && 71.419160 && 14.283${\pm}$0.035 && 13.749${\pm}$0.041 && 13.715${\pm}$0.052 && 13.606${\pm}$0.039 && 13.707${\pm}$0.058 && 0.85\\
6.753539  && 71.377299 && 14.392${\pm}$0.039 && 13.770${\pm}$0.043 && 13.590${\pm}$0.047 && 13.494${\pm}$0.032 && 13.511${\pm}$0.040 && 0.61\\
6.960778  && 71.371951 && 14.306${\pm}$0.036 && 13.917${\pm}$0.043 && 13.876${\pm}$0.061 && 13.693${\pm}$0.027 && 13.628${\pm}$0.035 && 0.55\\
6.776112  && 71.406968 && 14.423${\pm}$0.035 && 13.941${\pm}$0.042 && 13.812${\pm}$0.051 && 13.814${\pm}$0.041 && 13.834${\pm}$0.053 && 0.69\\
6.814866  && 71.462263 && 14.877${\pm}$0.043 && 14.374${\pm}$0.057 && 14.323${\pm}$0.079 && 14.058${\pm}$0.029 && 13.952${\pm}$0.042 && 0.51\\
6.853756  && 71.340099 && 14.792${\pm}$0.044 && 14.385${\pm}$0.059 && 14.210${\pm}$0.079 && 14.026${\pm}$0.038 && 14.010${\pm}$0.047 && 0.57\\
6.695908  && 71.426849 && 14.829${\pm}$0.047 && 14.387${\pm}$0.072 && 14.338${\pm}$0.078 && 14.203${\pm}$0.040 && 14.277${\pm}$0.057 && 0.60\\
6.711744  && 71.373390 && 14.777${\pm}$0.044 && 14.414${\pm}$0.060 && 14.333${\pm}$0.086 && 14.107${\pm}$0.033 && 14.116${\pm}$0.047 && 0.64\\
6.785655  && 71.390304 && 14.884${\pm}$0.046 && 14.416${\pm}$0.059 && 14.207${\pm}$0.069 && 14.342${\pm}$0.042 && 14.408${\pm}$0.059 && 0.87\\
6.698227  && 71.421794 && 14.881${\pm}$0.045 && 14.473${\pm}$0.060 && 14.412${\pm}$0.085 && 14.361${\pm}$0.039 && 14.439${\pm}$0.060 && 0.66\\
6.855808  && 71.429026 && 14.908${\pm}$0.044 && 14.513${\pm}$0.052 && 14.596${\pm}$0.095 && 14.418${\pm}$0.031 && 14.395${\pm}$0.050 && 0.80\\
6.910567  && 71.432888 && 14.928${\pm}$0.040 && 14.531${\pm}$0.063 && 14.488${\pm}$0.099 && 14.491${\pm}$0.054 && 14.513${\pm}$0.079 && 0.66\\
6.931771  && 71.421137 && 15.504${\pm}$0.068 && 14.601${\pm}$0.066 && 14.336${\pm}$0.087 && 14.323${\pm}$0.037 && 14.341${\pm}$0.055 && 0.76\\
7.021364  && 71.395864 && 15.112${\pm}$0.047 && 14.633${\pm}$0.068 && 14.495${\pm}$0.098 && 14.510${\pm}$0.041 && 14.652${\pm}$0.072 && 0.55\\
6.831960  && 71.358905 && 15.342${\pm}$0.059 && 14.727${\pm}$0.072 && 14.515${\pm}$0.086 && 14.352${\pm}$0.032 && 14.196${\pm}$0.048 && 0.67\\
 \hline
\end{tabular}
\end{center}
\end{table}
\newpage
\begin{table}
\caption{The $JHKW_1W_2$ catalogue of stars in \astrobj{DOLIDZE~14} having spatio-kinematic probability greater than 0.5.\label{tab3}}
\setlength{\tabcolsep}{2.5pt}
\tiny
\medskip
\begin{center}
 \begin{tabular}{ccccccccccccccccccccccccccc}
  \hline
RA&& DEC&& $J$ $\pm$ ${\sigma}_J$ &&$H$ $\pm$ ${\sigma}_H$ && $K$ $\pm$ ${\sigma}_K$ && $W_1$ $\pm$ ${\sigma}_{W_1}$ && $W_2$ $\pm$ ${\sigma}_{W_2}$ &&  $p_{sk}$\\
  \hline\noalign{\smallskip}
61.684179  && 27.443354 && 10.236${\pm}$0.022 && 9.595${\pm}$0.028 && 9.431${\pm}$0.021 && 9.335${\pm}$0.025 && 9.443${\pm}$0.020 && 0.60\\
61.561065  && 27.385290 && 11.322${\pm}$0.022 && 10.702${\pm}$0.030 && 10.587${\pm}$0.021 && 10.532${\pm}$0.023 && 10.572${\pm}$0.022 && 0.75\\
61.571746  && 27.415812 && 11.413${\pm}$0.022 && 10.994${\pm}$0.030 && 10.930${\pm}$0.021 && 10.848${\pm}$0.023 && 10.900${\pm}$0.020 && 0.74\\
61.612294  && 27.306736 && 11.911${\pm}$0.022 && 11.325${\pm}$0.030 && 11.207${\pm}$0.020 && 11.138${\pm}$0.026 && 11.199${\pm}$0.024 && 0.59\\
61.582959  && 27.434197 && 12.943${\pm}$0.024 && 12.312${\pm}$0.030 && 12.211${\pm}$0.026 && 12.157${\pm}$0.023 && 12.200${\pm}$0.024 && 0.56\\
61.591347  && 27.447421 && 12.883${\pm}$0.024 && 12.594${\pm}$0.031 && 12.515${\pm}$0.028 && 12.483${\pm}$0.025 && 12.482${\pm}$0.026 && 0.71\\
61.542977  && 27.363284 && 12.988${\pm}$0.022 && 12.776${\pm}$0.032 && 12.748${\pm}$0.026 && 12.670${\pm}$0.026 && 12.667${\pm}$0.029 && 0.74\\
61.664869  && 27.345145 && 13.089${\pm}$0.024 && 12.862${\pm}$0.031 && 12.749${\pm}$0.024 && 12.731${\pm}$0.027 && 12.701${\pm}$0.029 && 0.56\\
61.587348  && 27.384200 && 13.328${\pm}$0.024 && 12.891${\pm}$0.035 && 12.824${\pm}$0.032 && 12.769${\pm}$0.026 && 12.792${\pm}$0.029 && 0.85\\
61.681753  && 27.433113 && 13.264${\pm}$0.028 && 12.953${\pm}$0.038 && 12.843${\pm}$0.035 && 12.709${\pm}$0.027 && 12.721${\pm}$0.030 && 0.66\\
61.669820  && 27.315346 && 13.535${\pm}$0.026 && 13.134${\pm}$0.030 && 13.089${\pm}$0.026 && 12.999${\pm}$0.028 && 12.984${\pm}$0.031 && 0.66\\
61.486766  && 27.370216 && 13.385${\pm}$0.026 && 13.212${\pm}$0.033 && 13.226${\pm}$0.033 && 13.116${\pm}$0.026 && 13.122${\pm}$0.033 && 0.53\\
61.519334  && 27.314282 && 13.705${\pm}$0.027 && 13.215${\pm}$0.035 && 13.120${\pm}$0.034 && 13.003${\pm}$0.028 && 13.053${\pm}$0.032 && 0.59\\
61.549792  && 27.418307 && 13.924${\pm}$0.027 && 13.247${\pm}$0.032 && 13.108${\pm}$0.028 && 12.970${\pm}$0.025 && 12.947${\pm}$0.031 && 0.67\\
61.710244  && 27.307227 && 13.977${\pm}$0.028 && 13.617${\pm}$0.033 && 13.593${\pm}$0.032 && 13.501${\pm}$0.028 && 13.500${\pm}$0.042 && 0.53\\
61.535424  && 27.352599 && 13.925${\pm}$0.027 && 13.622${\pm}$0.033 && 13.539${\pm}$0.034 && 13.479${\pm}$0.028 && 13.471${\pm}$0.038 && 0.72\\
61.676643  && 27.378714 && 14.220${\pm}$0.030 && 13.778${\pm}$0.039 && 13.651${\pm}$0.041 && 13.524${\pm}$0.031 && 13.505${\pm}$0.040 && 0.76\\
61.592446  && 27.348249 && 14.158${\pm}$0.030 && 13.883${\pm}$0.031 && 13.804${\pm}$0.044 && 13.765${\pm}$0.029 && 13.771${\pm}$0.044 && 0.88\\
61.605692  && 27.418283 && 14.398${\pm}$0.028 && 13.921${\pm}$0.035 && 13.772${\pm}$0.044 && 13.725${\pm}$0.028 && 13.749${\pm}$0.044 && 0.68\\
61.531677  && 27.428993 && 14.251${\pm}$0.027 && 13.955${\pm}$0.043 && 13.839${\pm}$0.049 && 13.787${\pm}$0.029 && 13.875${\pm}$0.050 && 0.56\\
61.553499  && 27.415977 && 14.352${\pm}$0.034 && 14.014${\pm}$0.039 && 13.892${\pm}$0.047 && 13.859${\pm}$0.029 && 13.791${\pm}$0.047 && 0.69\\
61.551728  && 27.425931 && 14.449${\pm}$0.030 && 14.082${\pm}$0.043 && 14.050${\pm}$0.047 && 13.936${\pm}$0.031 && 13.918${\pm}$0.053 && 0.59\\
61.603401  && 27.342520 && 14.683${\pm}$0.035 && 14.248${\pm}$0.049 && 14.080${\pm}$0.046 && 13.949${\pm}$0.031 && 13.568${\pm}$0.040 && 0.69\\
61.611328  && 27.295861 && 14.571${\pm}$0.034 && 14.291${\pm}$0.046 && 14.294${\pm}$0.060 && 14.151${\pm}$0.032 && 14.079${\pm}$0.056 && 0.55\\
61.610973  && 27.352280 && 14.813${\pm}$0.043 && 14.385${\pm}$0.056 && 14.278${\pm}$0.066 && 14.015${\pm}$0.031 && 14.112${\pm}$0.054 && 0.93\\
61.580067  && 27.482747 && 14.810${\pm}$0.038 && 14.411${\pm}$0.055 && 14.310${\pm}$0.066 && 14.327${\pm}$0.031 && 14.333${\pm}$0.060 && 0.51\\
61.632290  && 27.419869 && 15.199${\pm}$0.040 && 14.425${\pm}$0.046 && 14.272${\pm}$0.060 && 14.180${\pm}$0.032 && 14.085${\pm}$0.055 && 0.65\\
61.718303  && 27.423905 && 15.221${\pm}$0.043 && 14.448${\pm}$0.053 && 14.468${\pm}$0.070 && 14.412${\pm}$0.036 && 14.426${\pm}$0.069 && 0.56\\
61.601538  && 27.333732 && 14.876${\pm}$0.041 && 14.471${\pm}$0.053 && 14.449${\pm}$0.064 && 14.347${\pm}$0.034 && 14.340${\pm}$0.062 && 0.83\\
61.667803  && 27.394607 && 14.891${\pm}$0.043 && 14.473${\pm}$0.056 && 14.465${\pm}$0.065 && 14.308${\pm}$0.033 && 14.214${\pm}$0.058 && 0.75\\
61.709741  && 27.375144 && 14.952${\pm}$0.043 && 14.515${\pm}$0.047 && 14.622${\pm}$0.084 && 14.460${\pm}$0.040 && 14.373${\pm}$0.067 && 0.64\\
61.539392  && 27.293732 && 15.143${\pm}$0.046 && 14.654${\pm}$0.058 && 14.370${\pm}$0.055 && 14.508${\pm}$0.035 && 14.453${\pm}$0.068 && 0.57\\
61.552529  && 27.368272 && 15.061${\pm}$0.038 && 14.654${\pm}$0.051 && 14.710${\pm}$0.087 && 14.510${\pm}$0.035 && 14.729${\pm}$0.088 && 0.72\\
61.515459  && 27.358919 && 15.369${\pm}$0.045 && 14.761${\pm}$0.056 && 14.848${\pm}$0.095 && 14.631${\pm}$0.037 && 14.536${\pm}$0.075 && 0.63\\
61.534298  && 27.431297 && 15.048${\pm}$0.045 && 14.761${\pm}$0.068 && 14.652${\pm}$0.079 && 14.541${\pm}$0.036 && 14.521${\pm}$0.076 && 0.56\\
61.554875  && 27.442931 && 15.242${\pm}$0.043 && 14.790${\pm}$0.067 && 14.886${\pm}$0.093 && 14.684${\pm}$0.038 && 14.663${\pm}$0.083 && 0.62\\
61.694496  && 27.311477 && 15.792${\pm}$0.067 && 14.945${\pm}$0.064 && 14.794${\pm}$0.093 && 14.704${\pm}$0.039 && 14.775${\pm}$0.093 && 0.58\\
61.697584  && 27.358894 && 15.681${\pm}$0.055 && 15.052${\pm}$0.069 && 14.902${\pm}$0.097 && 14.793${\pm}$0.042 && 14.772${\pm}$0.090 && 0.71\\
61.623794  && 27.296895 && 16.605${\pm}$0.127 && 16.084${\pm}$0.181 && 15.906${\pm}$0.000 && 16.248${\pm}$0.092 && 16.499${\pm}$0.408 && 0.61\\
 \hline
\end{tabular}
\end{center}
\end{table}
\begin{table}
\caption{Various physical parameters of the clusters under study. \label{tab4}}
\setlength{\tabcolsep}{1pt}
\small
\begin{center}
\begin{tabular}{cll}
\hline
Parameters & ~~~~\astrobj{NGC~110}~~~~ & ~~~~\astrobj{DOLIDZE~14}~~~~\\
\hline
RA~(J2000) & $00^{h}:27^{m}:22.4^{s}$ & $04^{h}:06^{m}:26.7^{s}$\\
DEC~(J2000) & $+71^{o}:23^{'}:56.6^{''}$ & $+27^{o}:22^{'}:26.6^{''}$\\
Core radius (arcmin) & $0.79{\pm}0.19$ & $2.76{\pm}0.81$\\
Cluster radius (arcmin)  & $5.6{\pm}0.4$ & $9.6{\pm}0.2$\\
~~Mean proper-motion (mas/yr)~~ & $4.03{\pm}0.29$, $2.53{\pm}0.23$ ~~~~& $-0.15{\pm}0.34$, $-7.79{\pm}0.41$~~~~\\
Log(age)~(yr) & $9.0{\pm}0.2$ & $9.1{\pm}0.2$\\
E(B-V) (mag) & $0.42{\pm}0.03$ & $0.32{\pm}0.02$\\
Distance-modulus (mag) & $10.57{\pm}0.28$ & $11.12{\pm}0.18$ \\
Distance (kpc) & $1.29{\pm}0.22$ & $1.67{\pm}0.14$\\
\hline
\end{tabular}
\end{center}
%
\end{table}
\end{document}